\documentclass[pre,twocolumn,twoside,byrevtex,superscriptaddress,floatfix]{revtex4-1}

\usepackage[realmainfile]{currfile}
\usepackage{url}
\usepackage{hyperref}
\hypersetup{
    colorlinks=true,
    linkcolor=grey,
    filecolor=grey,
    urlcolor=grey,
    breaklinks=true,
}

\usepackage[normalem]{ulem}

\usepackage[table]{xcolor}

\definecolor{olivegreen}{rgb}{0.33333,.41961,0.18431}
\definecolor{forestgreen}{rgb}{0.13333,.5451,0.13333}
\definecolor{lightgrey}{rgb}{0.7,0.7,0.7}
\definecolor{verylightgrey}{rgb}{0.90,0.90,0.90}
\definecolor{grey}{rgb}{0.5,0.5,0.5}

\definecolor{headerblue}{HTML}{33367E}
\definecolor{unitednationsblue}{HTML}{4D88FF}

\definecolor{charcoal}{HTML}{36454F}
\definecolor{cinerous}{HTML}{98817B}
\definecolor{feldgrau}{HTML}{4D5D53}
\definecolor{glaucous}{HTML}{6082B6}
\definecolor{arsenic}{HTML}{3B444B}
\definecolor{xanadu}{HTML}{738678}

\definecolor{firebrick}{HTML}{B22222}
\definecolor{orangered}{HTML}{FF4500}
\definecolor{tomato}{HTML}{FF6347}

\definecolor{purpletaupe}{HTML}{3B444B}

\definecolor{todoblue}{RGB}{0, 91, 187}

\usepackage{graphicx,epsfig,verbatim,enumerate}
\usepackage{amssymb,amsmath}
\usepackage{ifthen}

\usepackage{mathtools}

\newboolean{twocolswitch}

\newcommand{\sindex}[1]{}
\newcommand{\nindex}[1]{}

\newcommand{\www}[1]{\url{#1}}

\usepackage{lettrine}

\setboolean{twocolswitch}{true}

\begin{document}

\title{\protect
  Gauging the happiness benefit of US urban parks through Twitter
}

\author{
\firstname{Aaron J.}
\surname{Schwartz}
}

\email{aaron.j.schwartz@uvm.edu}

\affiliation{
  Gund Institute for Environment,
  University of Vermont,
  Burlington, VT 05405, USA
}

\affiliation{
  Vermont Complex Systems Center,
  University of Vermont,
  Burlington, VT 05401, USA
  }

\affiliation{
  Rubenstein School of Environment and Natural Resources,
  University of Vermont,
  Burlington, VT 05401, USA
}

\author{
\firstname{Peter Sheridan}
\surname{Dodds}
}

\affiliation{
  Vermont Complex Systems Center,
  University of Vermont,
  Burlington, VT 05401, USA
}

\affiliation{
  Computational Story Lab,
  MassMutual Center of Excellence for Complex Systems and Data Science,
  Vermont Advanced Computing Core,
  University of Vermont,
  Burlington, VT 05401, USA
}

\affiliation{
  Department of Mathematics \& Statistics,
  University of Vermont,
  Burlington, VT 05401, USA
  }

\author{
  \firstname{Jarlath P. M.}
  \surname{O'Neil-Dunne}
}

\affiliation{
  Gund Institute for Environment,
  University of Vermont,
  Burlington, VT 05405, USA
}

\affiliation{
  Vermont Complex Systems Center,
  University of Vermont,
  Burlington, VT 05401, USA
  }

\affiliation{
  Rubenstein School of Environment and Natural Resources,
  University of Vermont,
  Burlington, VT 05401, USA
}

\author{
  \firstname{Taylor H.}
  \surname{Ricketts}
}

\affiliation{
  Gund Institute for Environment,
  University of Vermont,
  Burlington, VT 05405, USA
}

\affiliation{
  Rubenstein School of Environment and Natural Resources,
  University of Vermont,
  Burlington, VT 05401, USA
}

\author{
  \firstname{Christopher M.}
  \surname{Danforth}
}

\affiliation{
  Vermont Complex Systems Center,
  University of Vermont,
  Burlington, VT 05401, USA
}

\affiliation{
  Computational Story Lab,
  MassMutual Center of Excellence for Complex Systems and Data Science,
  Vermont Advanced Computing Core,
  University of Vermont,
  Burlington, VT 05401, USA
  }

\affiliation{
  Department of Mathematics \& Statistics,
  University of Vermont,
  Burlington, VT 05401, USA
  }

\date{\today}

\begin{abstract}
  \protect
  The relationship between nature contact and mental well-being has
received increasing attention in recent years. While a body of
evidence has accumulated demonstrating a positive relationship between
time in nature and mental well-being, there have been few studies
comparing this relationship in different locations over long periods
of time. In this study, we estimate a happiness benefit, the
difference in expressed happiness between in- and out-of-park tweets,
for the 25 largest cities in the US by population. People write
happier words during park visits when compared with non-park user
tweets collected around the same time. While the words people write
are happier in parks on average and in most cities, we find
considerable variation across cities. Tweets are happier in parks at
all times of the day, week, and year, not just during the weekend or
summer vacation. Across all cities, we find that the happiness benefit
is highest in parks larger than 100 acres. Overall, our study suggests
the happiness benefit associated with park visitation is on par with
US holidays such as Thanksgiving and New Year's Day.
 
\end{abstract}


\maketitle

Human health and well-being depends on the environment in which we
live. Most people now live in cities, places defined by built
infrastructure where remnant nature and vegetation is planned or
managed. Urban greenspace, and specifically urban parks, can provide
opportunities to reduce the impacts of the ``urban health penalty,''
which includes higher levels of stress and depression in urban
residents (\cite{McDonald2018}). Nature contact is theorized to
promote mental health through several complementary pathways including
the physiological reduction of stress and the opportunity to restore
mental fatigue (\cite{Berto2014}). These pathways have been explored
using a dose-response framework which describe the duration,
frequency, and intensity of nature contact. Researchers have used
experimental, epidemiological, and experience-based approaches to
build a consensus around the mental health benefits of urban nature
(\cite{van2015health, krabbendam2020understanding}). However, there
are several questions remaining about how the benefits of nature
contact vary across cities (\cite{Frumkin2017a}).

Nature contact occurs within a specific context, and the ability of
urban residents to benefit from greenspace may vary
geographically. For example, four large cities showed different effect
sizes for their associations between nearby nature and well-being
(\cite{taylor2018wellbeing}). Larger studies have proved difficult
however; a recent review of studies on mental well-being and
greenspace in adults was unable to conduct a meta-analysis across
locations due to methodological heterogeneity
(\cite{houlden2018relationship}). Methods such as Ecological Momentary
Assessments and data from social media offer the opportunity to study
nature contact at wider spatial scales. Prior work using data from
Twitter has established that on average, in-park tweets are happier
than tweets originating elsewhere in cities
(\cite{Schwartz2019}). However, it has not been shown that this
pattern will hold across a wider selection of cities. The ability to
access and enjoy nature is heterogeneous across cities --- urban park
systems vary widely in quality and investment
(\cite{rigolon2018inequities}). A recent study found that county area
park expenditures were associated with better self-rated health
(\cite{mueller2019relationship}). We hypothesize that cities with
higher levels of investment in parks will provide greater benefits to
the mental well-being of park visitors. Understanding inter-city
variation in the mental health benefits of nature contact can inform
urban planning and public health policy.

The intensity of a dose of nature contact includes the size of the
natural area or park a person visits (\cite{Shanahan2015a,
  bratman2019nature}). Experimental approaches to nature contact are
limited in the number of natural areas they can integrate into their
study designs \cite{van2017urban}. A prior study found that the
visitors to the largest parks in San Francisco exhibited the greatest
mental benefits (\cite{Schwartz2019}). Here, we hypothesize that
larger parks will provide greater mental benefits to those who visit
them in cities, in general.

Studies using data from mobile phone applications and Twitter have
sampled over a time period between weeks and months and have not been
able to verify whether the timing (e.g., hour of day, day of week,
time of year) of park visits impacts potential health
benefits. However, a study using tweets in Melbourne demonstrated
heterogeneity in emotional responses to nature across different
seasons and time of day (\cite{Lim}). In addition, comparing the
benefits of park visitation temporally is a way to check the extent to
which observed happiness in parks is a function of park visits
occurring during the weekend or summer vacation.

Here, we expand our prior work in San Francisco to the 25 largest
cities in the US by population. For each city, we estimate a similar
metric of \textit{happiness benefit.} We compare this indicator across
cities using data from a four year period. We also compare the
happiness benefit across different categories of park size, as well as
across levels of city-wide park investment and quality. Finally, we
compare the happiness benefit of park visitation among different
seasons and days of the week.

\section{Material and Methods}
\label{methods}

\subsection{Data Collection \& Processing} 

We used a database of tweets collected from January 1 2012 to April 27
2015 (see Appendix \ref{appendix_twitter}), limiting our search to
English language tweets that included GPS coordinate location data
(latitude and longitude). We chose this time period because
geo-located tweets became abundant nationally in 2012 and dropped
significantly in April 2015 when Twitter made precise location sharing
an opt-in feature. Using boundaries from the US Census, we collected
tweets within each of the 25 largest cities in the US by population
(\cite{Census2012}). We did not include retweets (tweets that are
re-posted from another user) in our analysis.

We detected whether a tweet was posted within park boundaries using
the Trust for Public Land's Park Serve database. Our ability to find
tweets posted from inside parks depends on the accuracy of mobile GPS
hardware which can vary by manufacturer, surrounding building height,
and weather conditions. While most message locations should be precise
to within 10m, some of our user pool may have posted just outside of
parks due to measurement error. Data analysis of hashtag frequency
revealed that a large number of geo-located tweets were posted by
automated accounts (or bots) posting about job opportunities and
traffic; any tweet found with a job or traffic related hashtag was
removed from the sample (see Appendix \ref{appendix_hashtags}).

We assigned a control tweet to each in-park tweet. For each tweet, we
chose the closest-in-time out-of-park tweet from another user,
temporally proximate to the in-park tweet within the same city. This
message functions as a control because it allows us to compare the
happiness of our in-park sample with a set of tweets that were posted
in the same city and at roughly the same time.
We summarize each city's Twitter data in Tab.~\ref{tweet_table}.
In Appendix
\ref{appendix_user}, we describe an alternative control group
specification that uses out-of-park tweets from the same users who
posted tweets inside of parks.


\definecolor{lightgray}{gray}{0.9}
\rowcolors{1}{}{lightgray}

\def\arraystretch{1.5}%

\begin{table*}
\centering
\begin{tabular}{|l|c|c|c|c|c|c|} 
\hline
\textbf{\uline{City}} & \multicolumn{1}{l|}{\begin{tabular}[c]{@{}l@{}}\textbf{\uline{Total }}\\\textbf{\uline{tweets}}\end{tabular}} & \multicolumn{1}{l|}{\begin{tabular}[c]{@{}l@{}}\textbf{\uline{Park}}\\\textbf{\uline{tweets}}\end{tabular}} & \multicolumn{1}{l|}{\begin{tabular}[c]{@{}l@{}}\textbf{\uline{\% tweets }}\\\textbf{\uline{in parks}}\end{tabular}} & \multicolumn{1}{l|}{\begin{tabular}[c]{@{}l@{}}\textbf{\uline{Park }}\\\textbf{\uline{visitors}}\end{tabular}} & \multicolumn{1}{l|}{\begin{tabular}[c]{@{}l@{}}\textbf{\uline{Parks}}\\\textbf{\uline{visited}}\end{tabular}} & \multicolumn{1}{l|}{\begin{tabular}[c]{@{}l@{}}\textbf{\uline{Tweets }}\\\textbf{\uline{per capita}}\end{tabular}}  \\ 
\hline
New York      &    2,892,512 &     213,813 &               7.4 &       113,702 &         1,880 &               0.35 \\
Los Angeles   &    1,215,288 &      53,988 &               4.4 &        36,271 &           540 &               0.32 \\
Philadelphia  &    1,166,125 &      64,857 &               5.6 &        26,287 &           482 &               0.76 \\
Chicago       &    1,130,611 &      66,100 &               5.8 &        36,919 &           872 &               0.41 \\
Houston       &      821,433 &      39,581 &               4.8 &        13,464 &           501 &               0.38 \\
San Antonio   &      589,595 &      23,566 &               4.0 &        12,763 &           268 &               0.43 \\
Washington    &      570,157 &      74,937 &              13.1 &        41,062 &           370 &               0.92 \\
Boston        &      547,625 &      52,689 &               9.6 &        23,479 &           682 &               0.87 \\
San Diego     &      491,219 &      36,080 &               7.3 &        22,269 &           406 &               0.37 \\
Dallas        &      490,918 &      21,787 &               4.4 &        12,211 &           346 &               0.40 \\
San Francisco &      486,782 &      59,412 &              12.2 &        36,175 &           407 &               0.59 \\
Austin        &      449,853 &      23,547 &               5.2 &        14,689 &           289 &               0.55 \\
Baltimore     &      333,734 &      12,965 &               3.9 &         5,135 &           260 &               0.53 \\
Fort Worth    &      320,178 &       9,664 &               3.0 &         4,278 &           239 &               0.42 \\
Phoenix       &      268,455 &      12,041 &               4.5 &         7,566 &           189 &               0.18 \\
Columbus      &      251,573 &       8,884 &               3.5 &         4,340 &           328 &               0.31 \\
San Jose      &      234,234 &       8,263 &               3.5 &         4,517 &           314 &               0.24 \\
Indianapolis  &      225,931 &      11,560 &               5.1 &         5,660 &           183 &               0.27 \\
Charlotte     &      218,310 &       8,039 &               3.7 &         3,868 &           190 &               0.29 \\
Seattle       &      201,533 &      12,758 &               6.3 &         7,739 &           373 &               0.32 \\
Detroit       &      195,572 &       7,885 &               4.0 &         3,819 &           234 &               0.28 \\
Jacksonville  &      194,777 &       6,219 &               3.2 &         3,218 &           261 &               0.23 \\
Memphis       &      137,222 &       5,614 &               4.1 &         3,112 &           163 &               0.21 \\
Denver        &      131,240 &       6,243 &               4.8 &         3,902 &           279 &               0.21 \\
El Paso       &       96,015 &       2,722 &               2.8 &         1,397 &           180 &               0.14 \\
\hline
\end{tabular}
  \caption{
    Summary of geolocated Twitter data for the 25 most
    populous cities in the U.S. from 2012-2015.  `Total tweets'
    enumerates all public tweets posted from a GPS latitude/longitude
    inside that city. `Park tweets' is the total number of tweets
    posted from inside parks. The `\% tweets in park' column
    calculates Park tweets / total Tweets. `Park visitors' is the
    number of unique users who tweeted inside one of that city's
    municipal park locations as defined by Trust for Public Land's
    ParkServe.  `Parks visited' is the number of unique facilities
    from which a tweet was posted within that city.  `Tweets per
    capita' is number of total messages for the entire period divided
    by the city's population in 2012.  }
\label{tweet_table}
\end{table*}

\begin{figure*}[tp!]
    \centering
    \includegraphics[width=0.8\linewidth]{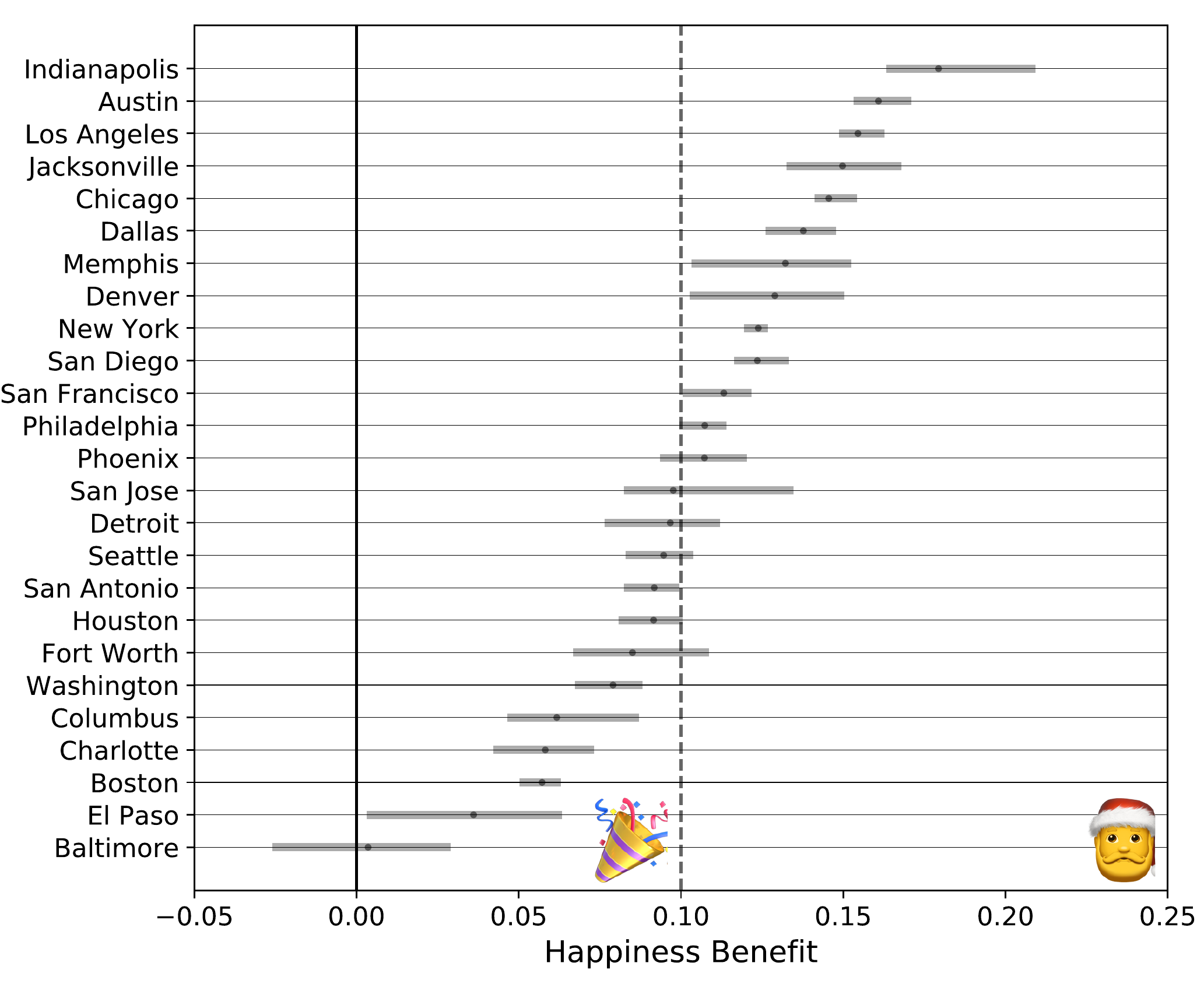}
    \caption{
      Happiness benefit by city, full range of values from 10
      bootstrap runs in which 80\% of tweets were randomly
      selected. The dark grey dots represent the mean value from
      bootstrap runs. For each city, the control group consists of
      non-park tweets posted at roughly the same time as each in-park
      tweet. The solid line marks a happiness benefit of 0, and the
      dotted line is average happiness benefit across all 25
      cities. Emojis denote the happiness benefit typically observed
      on New Year's Day and Christmas for all English tweets.
    }
    \label{fig_bumptime}
\end{figure*}

\subsection{Sentiment Analysis} 

To understand the mental benefits of park visitation, we used
sentiment analysis, a natural language processing technique that
associates numerical values to the emotional response induced by
individual words. For the present study, we used the Language
Assessment by Mechanical Turk (labMT) sentiment dictionary which
includes 10,222 of the most commonly used English words, merged from
four distinct text corpora, and rated on a scale of 1 (least happy) to
9 (most happy) (\cite{Dodds2011}). For example, \textit{beautiful} has
an average happiness score of $7.92$, \textit{city} has an average
happiness score of $5.76$, and \textit{garbage} has an average
happiness score of $3.18$ in labMT. We excluded words with scores
between $4.0$ and $6.0$ from our analysis because they are emotionally
neutral or particularly context dependent. The labMT sentiment
dictionary performs well when compared with other sentiment
dictionaries on large-scale texts, and correlates with traditional
surveys of well-being including Gallup's well-being index
(\cite{Reagan2017,Mitchell2013}). When using this type of bag-of-words
approach, it is inappropriate to rate the happiness levels of
individual tweets.

For each round of analysis, we aggregated tweets into an in-park group
and a control group. We calculated the average happiness for each
group of tweets as the weighted average of their labMT word scores
using relative word frequencies as weights:
\begin{equation}
  h_{\textnormal{avg}}
  =
  \frac{
    \sum_{i}^{N}
    h_{i}
    \cdot
    f_{i}
  }{
    \sum_i^Nf_{i}
  },
\end{equation} 
where $h_{i}$ is the happiness score of the ith word and $f_{i}$ is
its frequency in a group of tweets with $N$ words. Next, we subtracted
the average happiness of the control tweets from the average happiness
of the in-park tweets and defined this difference as the ``happiness
benefit''. To estimate uncertainty in our calculation of happiness
benefit, we applied a bootstrapping procedure: We randomly sampled
80\% of tweets without replacement from a set of in-park tweets and
their respective control tweets and then re-calculated the happiness
benefit. Performing this procedure 10 times, we derived a range of
plausible happiness benefit values. Robustness checks were performed
to show the convergence of this range at 10 runs.

We used the above technique to calculate the happiness benefit for all
cities together and each city individually. For each city, we removed
all words appearing in that city's park name before estimating the
happiness benefit. For example, we removed \textit{golden}, with an
average happiness of $7.3$, from all San Francisco tweets because of
Golden Gate Park. The word \textit{park} is also removed from all
tweets. We performed a manual check on the top ten most influential
words in a city's happiness benefit calculation. This allowed us to
identify potential biases introduced by words being used in an
unexpected manner. For example, we removed \textit{ma} from all Boston
tweets because it appears with a high frequency as an abbreviation for
Massachusetts, but has a positive happiness score as shorthand for
\textit{mother}. We include the full list of stop words in Appendix
\ref{appendix_stops}.

\subsection{Park Analysis}

We used data from the Trust for Public Land (TPL) to further
investigate the happiness benefit from urban park visits. The TPL
provides a variety of data on municipal park systems. Annually, TPL
publishes a ParkScore\textsuperscript{\tiny{\textregistered}} for the
largest cities in the US, which is a composite score out of 100 that
combines metrics of park size, access, investment, and amenities. We
conducted a correlation analysis for city-level happiness benefit
against 2018 ParkScore\textsuperscript{\tiny{\textregistered}} and
park spending per capita, also sourced from the TPL
(\cite{tpl2019}). ParkScore\textsuperscript{\tiny{\textregistered}}
and spending for Indianapolis was sourced from TPL's 2017 data release
due to lack of participation in 2018.

To investigate the relationship between happiness benefit and park
size, we assigned every in-park tweet a category based on the size of
the park from where it was posted. We grouped parks into four
categories ($<1$ acre, between $1$ and $10$ acres, between $10$ and
$100$ acres, and greater than $100$ acres). To have roughly equal
representation from each city, we randomly selected tweets (along with
their control tweet) in each park category from each city (or all of
the tweets in that category if there were less than 500). After
combining the randomly selected tweets from each city for each park
category, we estimated the happiness benefit using the same
bootstrapping procedure described above.

\subsection{Temporal Analysis}

Next, we estimate the happiness benefit based on when tweets were
posted in three different ways. First, we grouped tweets based on the
month they were posted in four seasonal groups (Winter: Dec, Jan, Feb;
Spring: Mar, Apr, May; Summer: Jun, Jul, Aug; Fall: Sep, Oct,
Nov). Second, we grouped tweets based on the day of the week they were
posted. Finally, we grouped tweets based on the hour of the day they
were posted in their local timezone (See Appendix
\ref{appendix_hour}). To have roughly equal representation from each
city, we randomly selected 1,000 tweets (along with their control
tweet) in each time category from each city (or all of the tweets in
that category if there were less than 1,000). After combining the
randomly selected tweets from each city, we estimated the happiness
benefit using the same bootstrapping procedure described above.

\section{Results}

\subsection{Sentiment Analysis} 

Across all cities, the mean happiness benefit was $0.10$ (Bootstrap
Range [$.098,.103$]). Across our 25 city sample, the mean happiness
benefit ranged from $0.00$ to $0.18$. Indianapolis had the highest
mean happiness benefit, while Baltimore had the lowest
(Fig.~\ref{fig_bumptime}). Cities with more in-park tweets to sample
from had tighter happiness benefit ranges, as exhibited by Denver, New
York, Los Angeles, and Philadelphia. The mean happiness benefit was
positive across all cities.

\subsection{Wordshifts}

\begin{figure}[tp!]
    \centering
    \includegraphics[width=\linewidth]{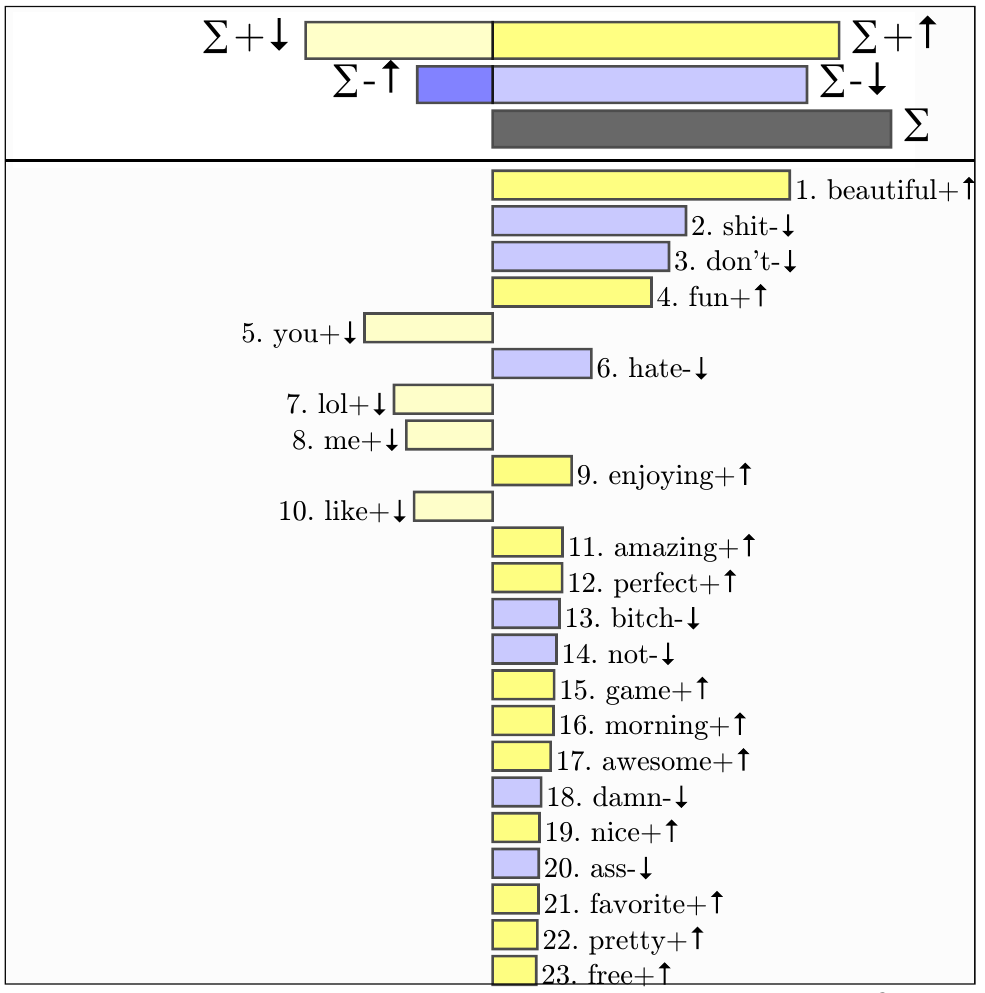}
    \caption{
      Differences in word frequency between park and control
      tweets across all cities, in order of decreasing contribution to
      the difference in average happiness. The right side represents
      the park tweets, with an average happiness of $5.96$. The left
      side represents the control tweets, with an average happiness of
      $5.86$. Purple bars represent words $\leq4$ (with $-$ symbol) on
      the Hedonometer scale. Yellow bars represent words $\geq6$ (with
      $+$ symbol) on the Hedonometer scale. Arrows indicate whether a
      word was more or less frequent within that set of tweets
      compared to the other text. For example, \textit{beautiful} is a
      positive word (yellow) with higher frequency in in-park tweets
      that contributes to its higher average happiness than the
      control tweets. \textit{Don't} is a negative word (purple) that
      appears less frequently in in-park tweets, also resulting in a
      higher average happiness score compared to control groups. Going
      against the overall trend, the positive words \textit{lol} and
      \textit{me} are used less often in parks. This wordshift uses
      tweets from 1,000 random in-park tweets and 1,000 control tweets
      from each city.
    }
    \label{fig_allcity}
\end{figure}

The happiness benefit is driven by word frequency differences between
the in-park tweets and control tweets. Specifically, positive words
(with a happiness score greater than $6$) including \textit{beautiful,
  fun, enjoying,} and \textit{amazing} appeared more frequently in
in-parks tweets. Negative words (with a happiness score less than $4$)
such as \textit{don't, not} and \textit{hate} appeared less frequently
in in-park tweets. We illustrate the variation in relative frequencies
in Fig.~\ref{fig_allcity}, a wordshift plot that demonstrates the most
influential words (by frequency and happiness) driving the happiness
benefit (\cite{Dodds2011}). Interactive versions of the city wordshift
graphs are available in the online appendix accompanying this
manuscript at
\url{http://compstorylab.org/cityparkhappiness/}.

\subsection{Park Analysis}

\begin{figure*}[tp!]
    \centering
    \includegraphics[width=0.8\linewidth]{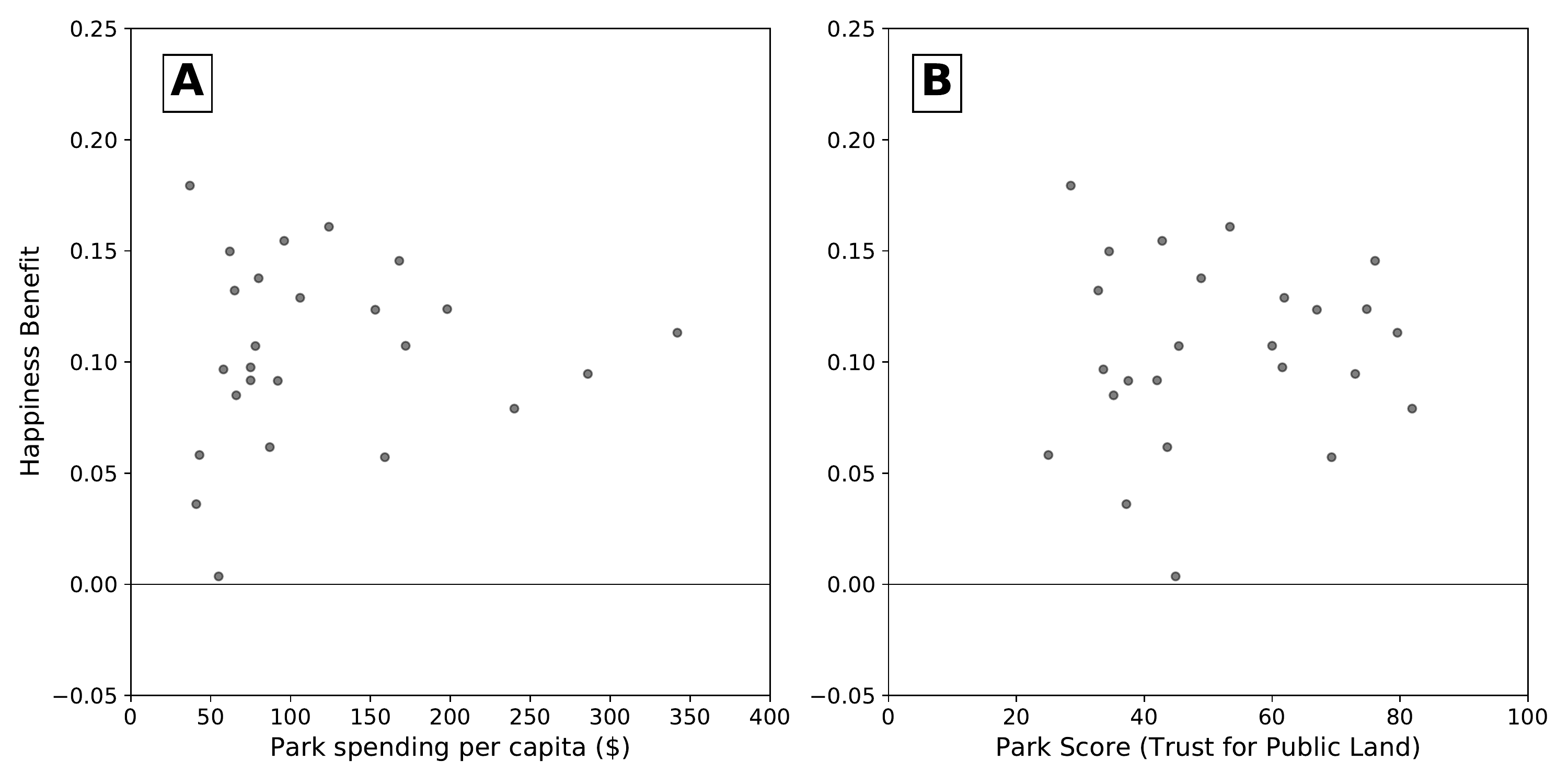}
    \caption{
      \textbf{A.}
      The left panel shows park spending per capita vs mean
      happiness benefit by city. Park spending per capita is from
      Trust for Public Land (TPL) data.
      \textbf{B.}
      The right panel shows
      ParkScore\textsuperscript{\tiny{\textregistered}} vs happiness
      benefit. The TPL calculates
      ParkScore\textsuperscript{\tiny{\textregistered}} annually from
      measures of park acreage, access, investment, and amenities, and
      is scaled to a maximum score of 100. The happiness benefit was
      not strongly correlated with per capita spending (Spearman's
      $\rho=0.14$) or
      ParkScore\textsuperscript{\tiny{\textregistered}} (Spearman's
      $\rho=0.03$).  }
    \label{fig_spendscore}
\end{figure*}

We plot the mean happiness benefit values against two metrics of park
quality --- park spending and
ParkScore\textsuperscript{\tiny{\textregistered}}
(Fig.~\ref{fig_spendscore}). There is no clear pattern between
happiness benefit and park spending or
ParkScore\textsuperscript{\tiny{\textregistered}}. Interestingly,
Indianapolis, which had the highest mean happiness benefit, had the
lowest municipal park spending per capita and one of the lowest
ParkScore\textsuperscript{\tiny{\textregistered}} values. Washington
D.C., San Francisco, Chicago, New York, and Seattle had the highest
ParkScore\textsuperscript{\tiny{\textregistered}} values, and were all
fairly close to the mean happiness benefit of $0.10$.

\begin{figure*}[tp!]
    \centering
    \includegraphics[width=\linewidth]{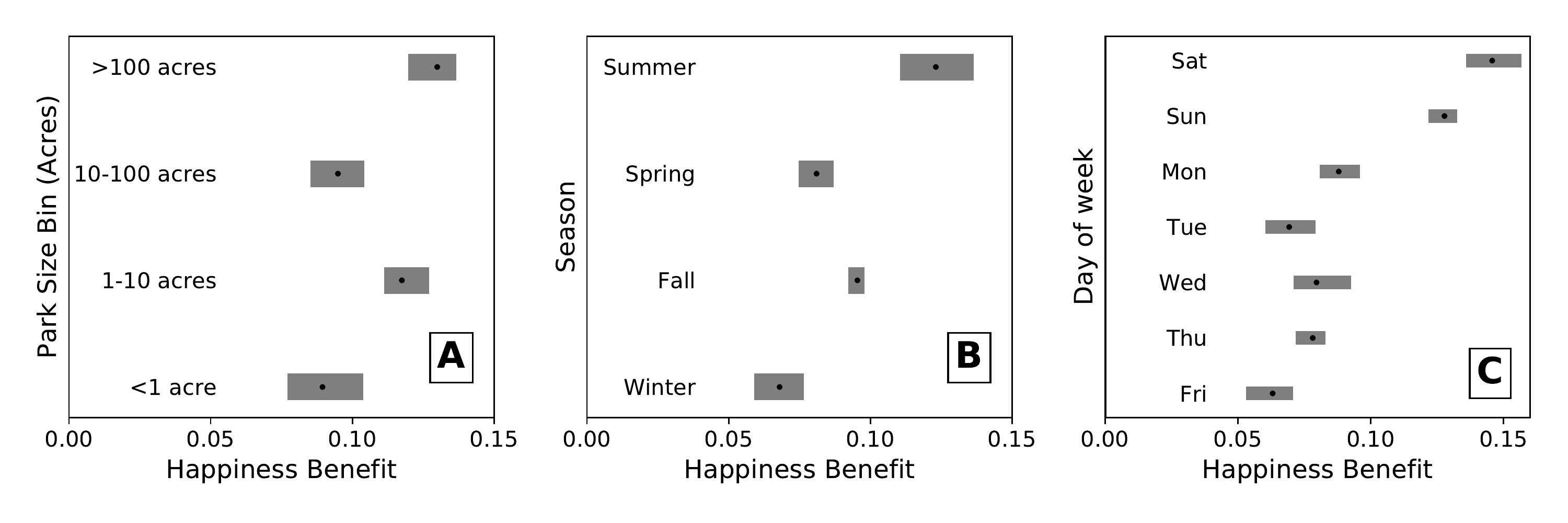}
    \caption{
      \textbf{A.} The left panel shows happiness benefit by park size. The
      largest category of parks (greater than 100 acres) had the
      highest happiness benefit.
      \textbf{B.} The middle panel shows happiness
      benefit by season, with summer and fall exhibiting the highest
      mean happiness benefit values.
      \textbf{C.} The right panel shows
      happiness benefit by day of the week, with the weekend days
      higher than other days of the week. In all three panels, the
      range is the full range of happiness benefits from 10 runs,
      sampling 80\% of tweets. 1,000 random in-park tweets were pooled
      in each group from each city. Control tweets were selected as
      tweets most temporally proximate to the in-park tweet from the
      same city.
    }
    \label{fig_bins}
\end{figure*}

We grouped in-park tweets into four categories based on the size of
the park and estimated the happiness benefit for each category. Parks
greater than $100$ acres had the highest mean happiness benefit of
$0.13$, followed by parks from $1-10$ acres ($0.12$). Parks less than
$1$ acre and parks between $10-100$ acres had the lowest mean
happiness benefit of $0.09$ (Fig.~\ref{fig_bins}).

\subsection{Comparing Cities}

\begin{figure*}[tp!]
  \centering
  \includegraphics[width=0.8\textwidth]{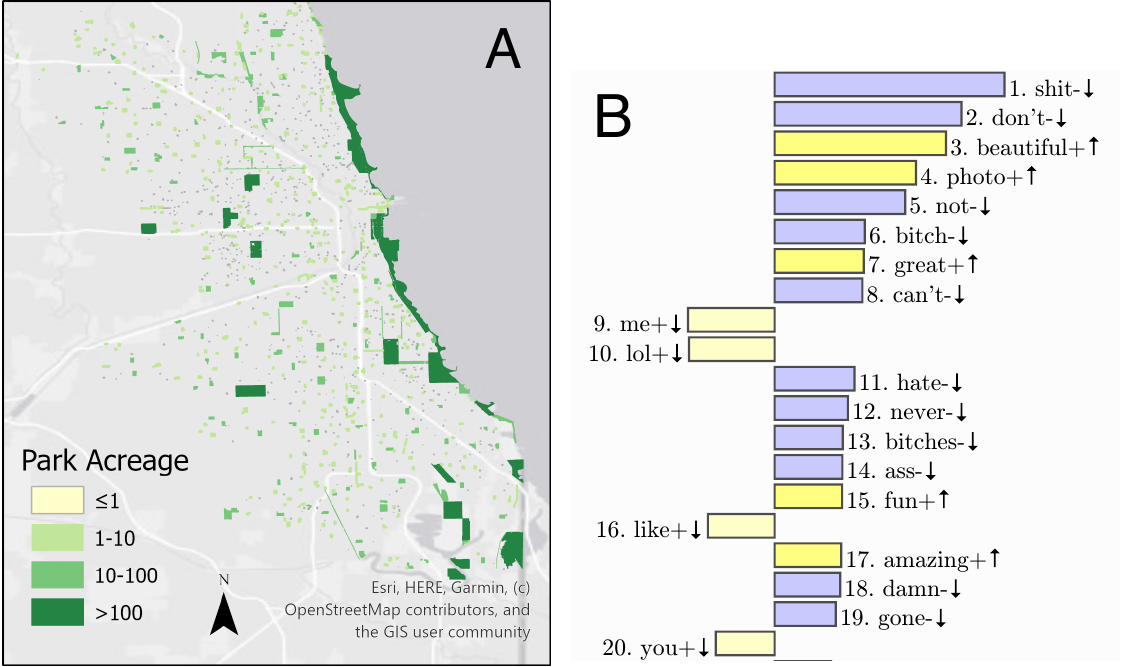}
    \caption{
      \textbf{A.} The left panel shows a map of the greater Chicago area and
      its municipal parks, shaded by park size category.
      \textbf{B.} The right
      panel is a wordshift for Chicago's tweets. In this wordshift the
      right side represents the park tweets. The left side represents
      the control tweets. Purple bars represent words $\leq4$ (with
      $-$ symbol) on the Hedonometer scale. Yellow bars represent
      words $\geq6$ (with $+$ symbol) on the Hedonometer scale. Arrows
      indicate whether a word was more or less frequent within that
      set of tweets compared to the other text. Individuals use
      positive words such as \textit{beautiful} and \textit{fun} more
      often in Chicago parks, and use profanity less often.
    }
    \label{fig_chicago}
\end{figure*}

We analyzed the average happiness of each individual city's
tweets. For example, Chicago had over 1.1 million total tweets, with
36,919 users tweeting from a park. Tweets were posted in 872 separate
park units, second only to New York. Roughly 6\% of all Chicago tweets
were posted from a park, with .41 tweets per capita from
2012--2015. Chicago's happiness benefit was $0.15$, ranking fifth
among our 25 cities. Chicago's tweets were distributed among many
different types of parks, including several large parks along the
shore of Lake Michigan. Tweets posted in Chicago parks had higher
average happiness than tweets posted elsewhere in Chicago due to
higher frequency of happy words such \textit{beautiful} and
\textit{great}, and lower frequency of unhappy words including
profanity, \textit{don't}, and \textit{not}. We include a map of
Chicago's parks and a wordshift plot between Chicago's in-park and
control tweets in Fig.~\ref{fig_chicago}.

\subsection{Temporal Analysis}

Across all cities, we grouped park tweets and their control tweets
according to the in-park tweet's timestamp. First, we compared the
happiness benefit by season. The mean happiness benefit was highest in
the summer ($0.12$), followed by fall ($0.10$), spring ($0.08$), and
winter ($0.06$) as shown in Fig.~\ref{fig_bins}. Then we grouped park
tweets and their respective control tweets according to the day of the
week in which it was posted. Saturday exhibited the highest mean
happiness benefit ($.15$) followed by Sunday ($0.13$). Monday through
Friday were all between $0.06$ and $0.09$ (Fig.~\ref{fig_bins}). We
also estimated the happiness benefit by hour of the day in which the
in-park tweet was posted. The tweets posted during the 8:00 and 9:00
AM hours had a mean happiness benefit around $0.07$ while the rest of
the day did not show a clear pattern, ranging from $0.08$ to $0.14$
(see Appendix \ref{appendix_hour}).

\section{Discussion}

\subsection{Sentiment Analysis} 

In this study, across the 25 largest cities in the US, we find that
people write happier words on Twitter in parks than they do outside of
parks. This effect is strongest for the largest parks by area -
greater than 100 acres. The effect is present during all seasons and
days of the week, but is most prominent during the summer and on
weekend days.

Pooling tweets across cities, we find a mean happiness benefit of
$0.10$. According to Hedonometer.org, which tracks Twitter happiness
as a whole using the labMT dictionary, Twitter has fluctuated around a
mean happiness of $6.02$ since 2008. New Year's Day has historically
had an average happiness of $6.11$, giving it an average happiness
benefit of $.10$. Christmas, historically the happiest day of the year
on Twitter, has had an average happiness benefit of $0.24$. The global
COVID-19 Pandemic gained rapid recognition in the US on March 12,
2020, which resulted in the then unhappiest day in Twitter's history
with a drop of $0.31$ from its historical average. Following the
murder of George Floyd, the Black Lives Matter protests led to a new
all-time low, $0.39$ below the historical average
(\cite{Hedonometer2020}). These are considered large swings, and we
assess that the happiness benefit of $0.10$ across a sample of 25,000
tweets is a strong signal.

Positive words such as \textit{beautiful, fun}, and \textit{enjoying}
contributed to the higher levels of happiness from our in-park tweet
group. These words may relate to the stimulating aspects of urban
greenspace. This is supported by a recent study that analyzed tweets
to investigate which aspects of restoration were most prominent in
urban greenspace. They found that fascination, an emotional state
induced through inherently interesting stimuli, was most salient
(\cite{wilkie2020attention}). Fascination is one characteristic of
nature experiences described by Attention Restoration Theory, which
theorizes that time in nature provides an opportunity to recover from
the cognitive fatigue induced by mentally taxing urban environments
(\cite{kaplan1989experience, Kaplan1995}).

We find high levels of variation across cities for the happiness
benefit between in-park and out-of-park tweets. In Chicago, higher
frequencies of words such as \textit{beautiful} drive higher in-park
tweet happiness. Park tweets had lower frequencies of negative words
such as \textit{don't, not}, and \textit{hate}
(Fig.~\ref{fig_chicago}). Psychological experiments treat positive and
negative affect as separate measures (\cite{mcmahan2015effect}); the
heterogeneity of the words driving the happiness benefit may be
related to how these components of affect are being expressed via
tweets.

\subsection{Park Analysis}

Park spending per capita and
ParkScore\textsuperscript{\tiny{\textregistered}} were not correlated
with mean happiness benefit by city. However, prior work has
demonstrated an association between park investment and levels of
self-rated health (\cite{mueller2019relationship}). Another study
found higher levels of physical activity and health to be associated
with a composite score of park quality in $59$ cities
(\cite{mullenbach2018peer}). Other factors such as heterogeneous use
patterns of Twitter across cities may be more associated with
happiness benefit than measures of park quality and spending. We
encourage further investigation of the relationship between park
quality and investment with the mental health benefits of nature
contact.

Tweets inside of all park size categories exhibited a positive
happiness benefit. The largest parks, greater than 100 acres, had the
highest mean happiness benefit. One possible explanation is that
larger parks provide greater opportunities for mental restoration and
separation from the taxing environment of the city. This finding is
consistent with results from our earlier study in San Francisco, in
which tweets in the larger and greener Regional Parks had the highest
happiness benefit (\cite{Schwartz2019}). Parks between 0 and 10 acres
are often neighborhood parks that people use in their day to day
lives. Local parks provide many essential functions; however, our
results suggest that the experiences people have in larger parks may
be more beneficial from a mental health perspective. Another
possibility is that people spend more time in larger parks; one study
suggested that 120 minutes of nature contact a week resulted in
improved health and well-being (\cite{White2019}).

\subsection{Temporal Analysis}

We observe that the mean happiness benefit was higher in summer than
other seasons; however, the happiness benefit was positive in all four
seasons. Similarly, the mean happiness benefit was highest during the
weekend, but positive on all days of the week (See
Fig.~\ref{fig_bins}). People use happier words when visiting parks
throughout the week and year --- not just outside of typical working
hours. This result is encouraging because some prior studies on nature
contact using Twitter only addressed shorter time spans. Future
studies should seek methods that can investigate the other temporal
aspects of nature contact including the frequency and duration of
visits (\cite{Shanahan2015a}).

\section{Concluding Remarks}

Future research should continue to explore the relationship between
tweet happiness and other factors beyond park investment.

While
ParkScore\textsuperscript{\tiny{\textregistered}} captures a variety
of park-quality related metrics, vegetation and biodiversity are
salient features of greenspace that significantly impact how people
experience their time in nature
(\cite{mavoa2019higher,wang2019makes,clark2014biodiversity}).
More
localized studies could look at the mental health impact of park-level
vegetative cover and biodiversity metrics.

While we investigated the
seasonal variation of in-park happiness, climate and weather have been
shown to influence happiness on Twitter as well
(\cite{baylis2018weather,moore2019rapidly}).
Tweets could be binned by
some composite of temperature, humidity, and precipitation in order to
investigate how weather moderates the association between nature
contact and mental well-being (\cite{van2017urban}).

Some greenspaces
are more crime prone than others and a recent study was able to
identify crime-related tweets, which may help further explain
happiness differences between parks
(\cite{kimpton2017greenspace,curiel2020crime}).
Demographic,
socioeconomic, and cultural factors also play a role in how people
engage with parks (\cite{browning2018income}).

While identifying such
factors on Twitter is challenging and requires ethical consideration,
other methodologies can continue to explore how different groups use
and benefit from time in parks, to help ensure that the benefits of
parks are available to everyone.
As the evidence continues to mount on
the many different benefits of nature contact, ensuring access to
quality parks for all urban residents is critical.

\bigskip

\acknowledgments
We are grateful for support from the NSF GRFP program, the Gund
Institute for the Environment Catalyst Award program, and a gift from
MassMutual Life Insurance Company.

\clearpage

\renewcommand{\thefigure}{A\arabic{figure}}
\setcounter{figure}{0}

\renewcommand{\thetable}{A\arabic{table}}
\setcounter{table}{0}

\renewcommand{\thesection}{A\arabic{section}}
\setcounter{section}{0}

\section{Appendix}

\subsection{Twitter API}
\label{appendix_twitter}

Twitter's `spritzer' streaming API offers a random selection of up to
1\% of all messages, with specific linguistic or spatial filters
enabling a higher percentage. For the present study, we collected
messages tagged with GPS coordinates during the years
2012--2015. During this period, geolocated messages comprised roughly
1\% of all messages. As a result, filtering on GPS enabled us to
collect nearly 100\% of all such messages.

\subsection{Stopwords}
\label{appendix_stops}

As is common in natural language processing, we define `stop words' as
individual words that we mask from sentiment analysis. These are words
that we identify as frequent in our tweets, but that contribute
neutral or context-dependent sentiment. We do not include the word
\textit{park} in our analysis. We removed the words \textit{closed,
  traffic,} and \textit{accident} because they frequently appeared in
geo-located tweets from automated traffic posts.  We removed words
found in the names of the parks (e.g., \textit{golden} and
\textit{gate}). Several cities had increased frequencies for the
positive words \textit{art, museums, gardens,} and \textit{zoos} in
their parks. Even though these words were not in the official park
names, we removed them from our analysis. Several parks had the
positive words \textit{music} and \textit{festival} appear frequently,
so we removed these two words. For each city, we identified a list of
stop words to remove by manually checking the 10 most influential
words contributing to the difference between in-park and control
tweets. Finally, we removed words that referred to a specific location
(e.g., \textit{beach}) or were being used in a significantly different
way than they were originally rated for happiness (e.g. \textit{ma} as
shorthand for Massachusetts rather than mother) were removed (See
Tab. \ref{word_table}).

\begin{figure}[h]
    \centering
    \includegraphics[width=\linewidth,keepaspectratio]{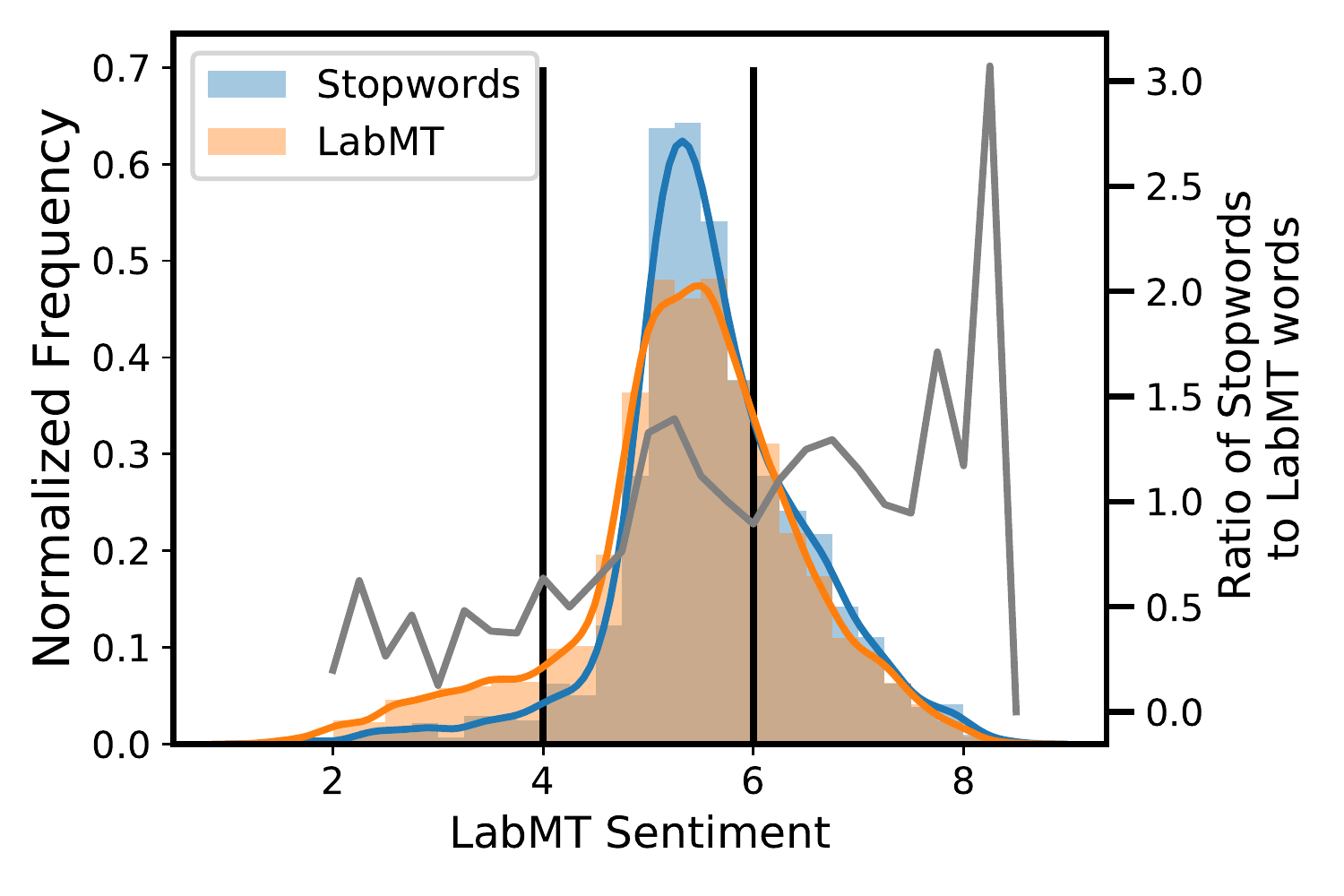}
    \caption{
      Normalized histogram of LabMT words and stop words taken out of
      the analysis due to being in a park name. Our analysis is
      conservative as the ratio is higher for positive words ($>6$)
      compared to negative words ($<4$). Words between $4$ and $6$ are
      not included in our analysis.
    }
    \label{fig_stophist}
\end{figure}

Overall, the majority of words we masked were positive, with average
happiness scores greater than 6 as seen in Fig. \ref{fig_stophist}. As
a result, we expect that the happiness benefit reported in our results
is a lower bound.

\subsection{Hashtags}
\label{appendix_hashtags}

Tweets with any of the following hashtags were removed from our study
sample:
\#jobs,
\#job,
\#getalljobs,
\#hiring,
\#tweetmyjobs,
\#careerarc,
\#hospitality,
\#healthcare,
\#nursing,
\#marketing,
\#sales,
\#clerical,
and
\#it.

\subsection{Happiness Benefit: User Control}

\begin{figure*}[h]
    \centering
    \includegraphics[width=0.8\linewidth]{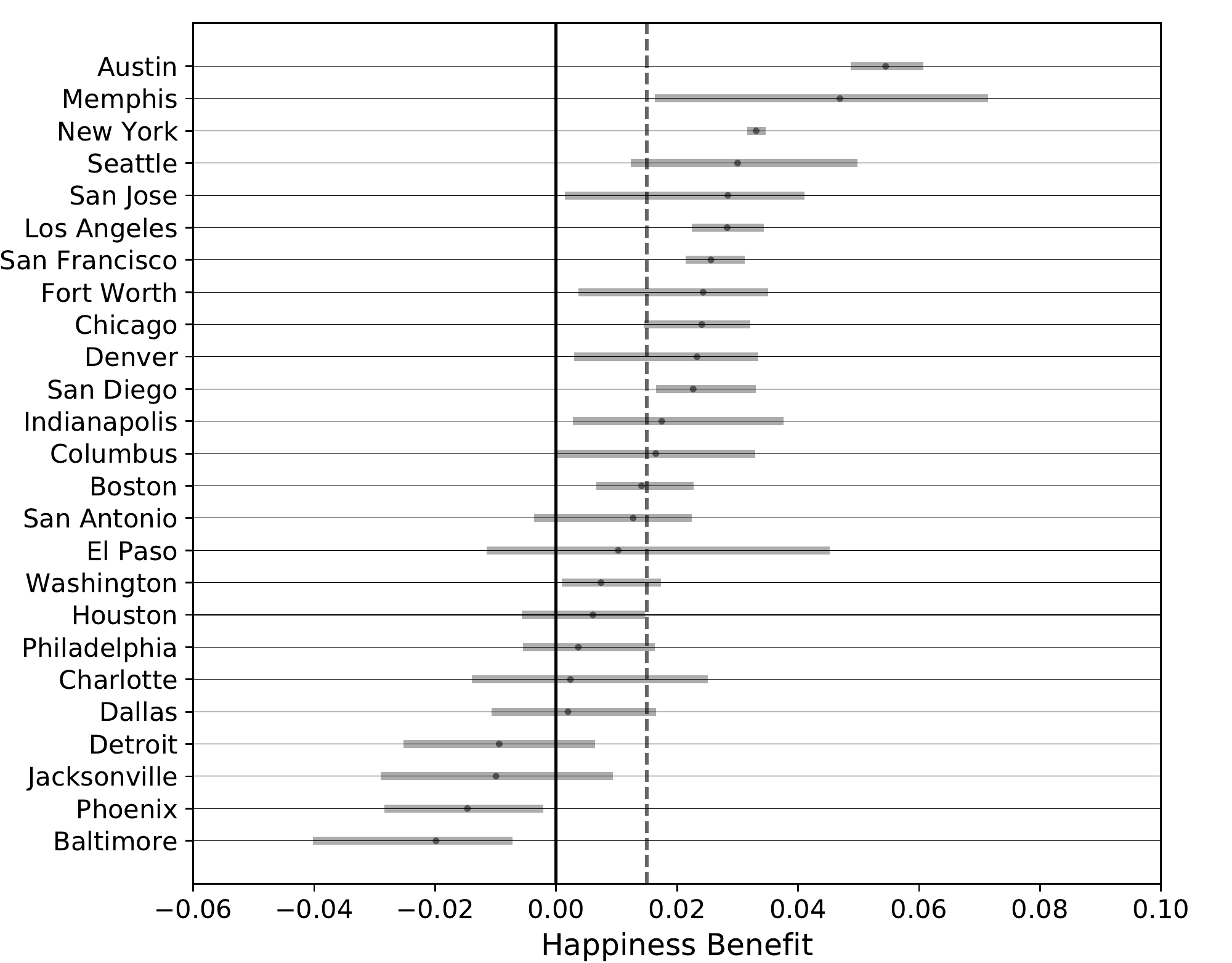}
    \caption{
      Happiness benefit by city. We derive each city's full range of
      values from 10 bootstrap runs, for which we randomly selected
      80\% of tweets. Darker dots represent mean value from bootstrap
      runs. For each city, the control group consists of 1 random,
      non-park tweet from each user paired with an in-park tweet.
    }
    \label{fig_bumpuser}
\end{figure*}

\begin{figure*}[h]
  \centering
  \includegraphics[width=0.8\linewidth]{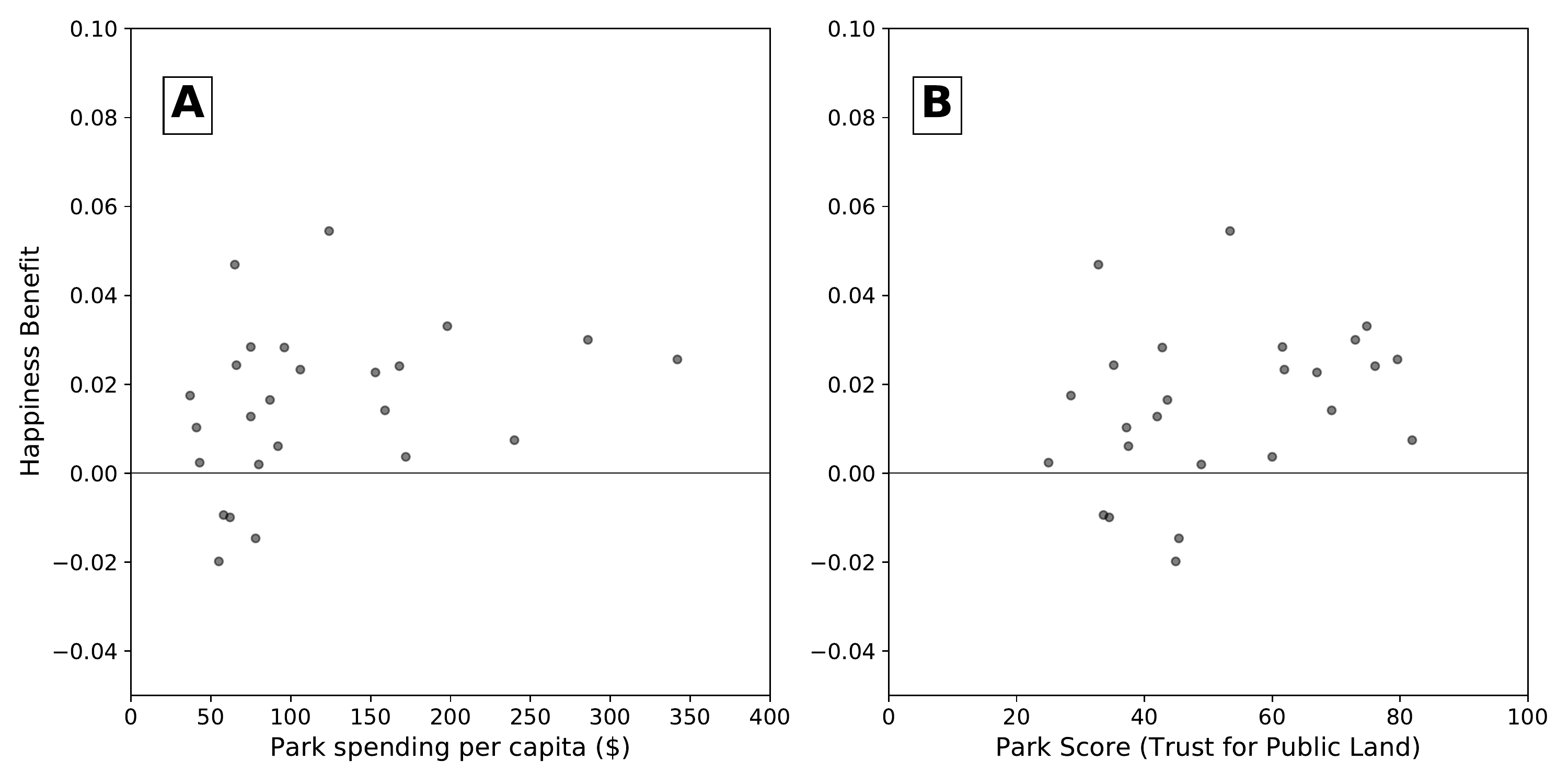}
  \caption{
    A. The left panel shows park spending per capita vs mean
    happiness benefit by city. Park spending per capita is from
    Trust for Public Land (TPL) data. B. The right panel shows
    ParkScore\textsuperscript{\tiny{\textregistered}} vs mean
    happiness. The TPL calculates
    ParkScore\textsuperscript{\tiny{\textregistered}} annually from
    measures of park acreage, access, investment, and amenities, and
    is scaled to a maximum score of 100.
  }
  \label{fig_spendscore_user}
\end{figure*}

In addition to the proximate time control described in the Methods
section, we employed a secondary control to investigate the happiness
benefit methodology. In this method, we selected a `user control'
tweet: a random message from the same user posted out-of-park. If an
account's message history consisted entirely of in-park tweets, the
account was removed from the sample as they were likely a tourist or
business located adjacent to the park. The user control allows us to
estimate a happiness benefit for the users during their park visits
compared to tweets when they were not in the parks. We performed the
same happiness benefit calculation for each of the 25 cities and
include those results in Figure \ref{fig_bumpuser}. For our `user
control' group, the mean happiness benefit for the cities in our
sample ranged from $-0.02$ to $.05$ (Fig. \ref{fig_bumpuser}). We also
plot the mean happiness benefit against park spending for capita and
Park Score\textsuperscript{\tiny{\textregistered}} in
Fig. \ref{fig_spendscore_user}. The overall benefit reduction observed
for the User Control, when compared with the time control, suggests
that individuals who tweet from within parks generally use happier
words than individuals who do not visit parks.
\label{appendix_user}

\subsection{Temporal Analysis by hour of day}
\label{appendix_hour}

\begin{figure*}[ht]
    \centering
    \includegraphics[width=0.8\linewidth]{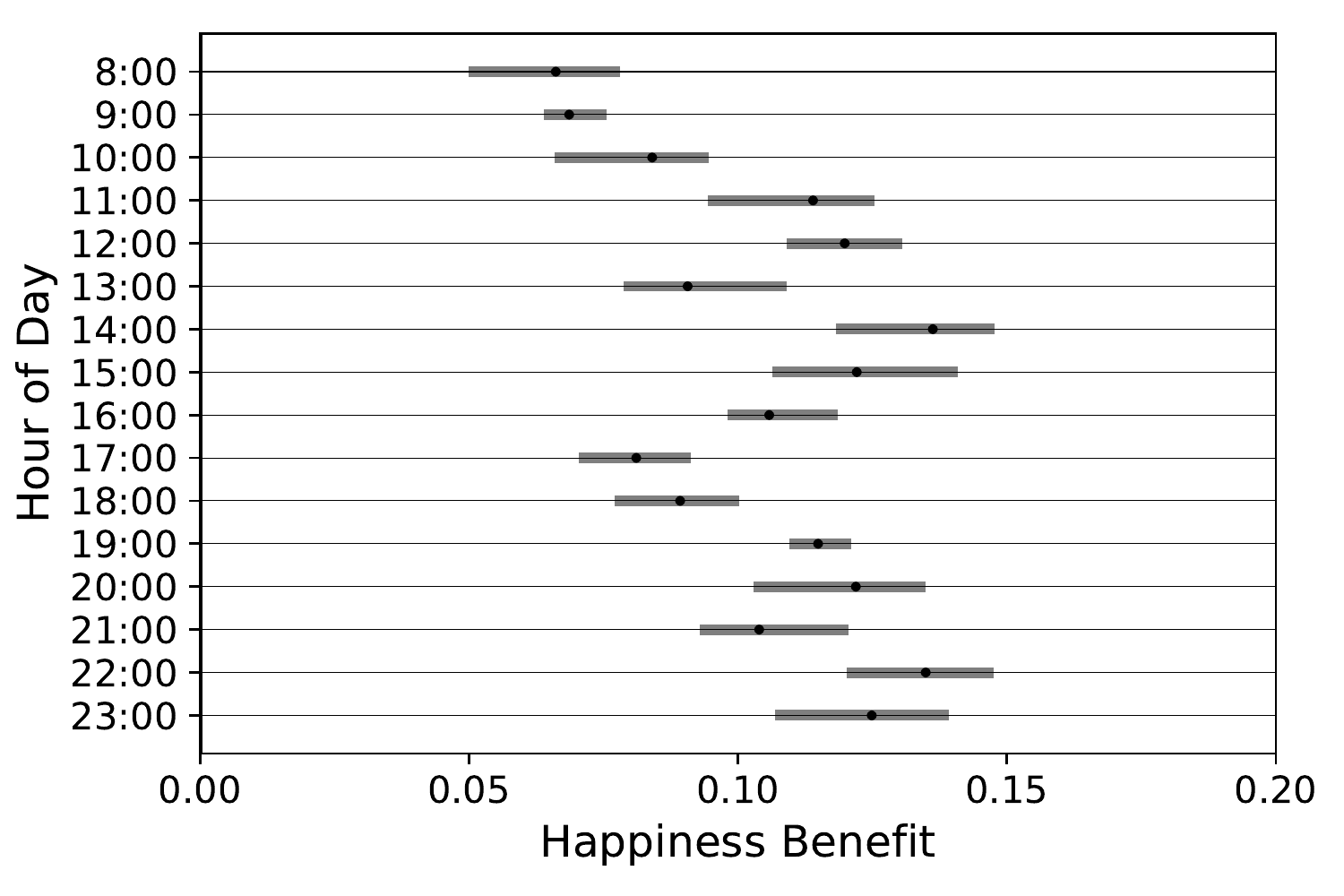}
    \caption{
      Change in happiness benefit by hour of day. The range is the
      full range of happiness benefit estimates from 10 runs, sampling
      80\% of tweets. 1,000 random in-park tweets were pooled in each
      group from each city. Control tweets were selected as tweets
      most temporally proximate to the in-park tweet from the same
      city.
    }
    \label{fig_hour}
\end{figure*}

We estimated the happiness benefit by hour of day across all cities
(Fig. \ref{fig_hour}). While 8:00 and 9:00AM are slightly lower, the
rest of the day's happiness benefit ranges overlap, showing that our
other results are not biased by certain hours of the day (e.g.,
leaving the office).


\definecolor{lightgray}{gray}{0.9}
\rowcolors{1}{}{lightgray}

\def\arraystretch{1.5}%

\begin{table*}
\centering
\begin{tabular}{|l|l|} 
\hline
\textbf{\uline{City }} & \textbf{\uline{Stop Words }}    \\ 
\hline
San Francisco          & young, flowers                  \\ 
\hline
Phoenix                & hospital                        \\ 
\hline
Jacksonville           & science                         \\ 
\hline
Austin                 & limits                          \\ 
\hline
San Diego              & sea                             \\ 
\hline
Washington             & war, bill, united, health       \\ 
\hline
Seattle                & health, surgery, emergency      \\ 
\hline
Chicago                & riot                            \\ 
\hline
Houston                & hospital, delay, stop, science  \\ 
\hline
Cleveland              & beach, island                   \\ 
\hline
Boston                 & ma, partners                    \\ 
\hline
New York               & natural                         \\ 
\hline
San Antonio            & cafe                            \\ 
\hline
Dallas                 & health                          \\ 
\hline
Philadelphia           & independence                    \\ 
\hline
Los Angeles            & science                         \\ 
\hline
San Jose               & christmas, raging               \\ 
\hline
Denver                 & nature, science, international  \\ 
\hline
Memphis                & steal, sugar                    \\ 
\hline
Charlotte              & shot, young                     \\ 
\hline
Indianapolis           & health                          \\ 
\hline
Columbus               & roses                           \\
\hline
\end{tabular}
\caption{Stop words selected for individual cities based on frequency analysis and contextual meaning.}
\label{word_table}
\end{table*}

\end{document}